\documentclass[manuscript]{aastex}

\usepackage{amsmath}
\numberwithin{equation}{section}

\slugcomment{}

\begin{document}


\title{A 77-118 GHz Resonance-Free Septum Polarizer}


\author{Yen-Lin Chen\altaffilmark{1} }
\affil{National Taiwan University Department of Physics}
\author{Tzihong Chiueh\altaffilmark{1,2,3}}
\affil{National Taiwan University Department of Physics}
\author{Hsiao-Feng Teng\altaffilmark{1}}
\affil{National Taiwan University Department of Physics}
\email{chiuehth@phys.ntu.edu.tw}


\altaffiltext{1}{Department of Physics, National Taiwan University, 106. Taipei, Taiwan}
\altaffiltext{2}{Institute of Astrophysics, National Taiwan University, 106. Taipei, Taiwan}
\altaffiltext{3}{Center for Theoretical Sciences, National Taiwan University, 106. Taipei, Taiwan}


\begin{abstract}
Measurements of the polarized radiation often reveal specific physical properties of emission sources, such as  strengths and orientations of magnetic fields offered by synchrotron radiation and Zeeman line emission, and the electron density distribution by free-free emission. Polarization-capable, millimeter/sub-millimeter telescopes are normally equipped with either septum polarizers or ortho-mode transducers (OMT) for the detection of polarized radiation.  While the septum polarizer is traditionally conceived to be limited to a significantly narrower bandwidth than the OMT, it does possess advantageous features for astronomical polarization measurements unparalleled by the OMT.  Challenging the conventional bandwidth limit, we design an extremely wideband circular waveguide septum polarizer, covering $42\%$ bandwidth, from 77 GHz to 118 GHz, without any undesired resonance.  Stokes parameters constructed from the measured data in between 77 GHz and 115 GHz show that the leakage from $I$ to $Q$ and $U$ is below $\pm 2\%$ and the $Q-U$ mutual leakage below $\pm 1\%$.  Such a performance is comparable to other modern polarizers, but the bandwidth of this polarizer can be at least twice as wide.   This extremely wide-band design removes the major weakness of the septum polarizer and opens up a new window for future astronomical polarization measurements.
\end{abstract}





\section{Introduction}

Wide-band polarization measurements in millimeter and sub-millimeter Astronomy have always been a challenge.   Examples requiring wide-bandwidth polarimetry measurements include the continuum cosmic microwave background (CMB) polarization observation \citep{kov02, kog03, bar05, mon06, sie07, fri09, chi10, bis11}, the synchrotron polarization observations of compact sources \citep{dow98, ait00, cul11}, the polarization observation of thermal dust emission \citep{laz03, bet07, hoa11} and the observation of Zeeman effects via molecular lines, such as CN and SO \citep{bel89, cru99, shi00}. Except for some particular features of Zeeman effects, most of the above focus on linear polarization measurements.  For the continuum observation, it is desirable the available frequency bandwidth to be as wide as possible so that, on one hand, the signal-to-noise ratio can increase, and on the other hand the spectral index may be determined. For the line observation, the emission line can be red-shifted to any unknown frequency when emitted from, or absorbed at, a distant universe, and hence wide frequency coverage has a unique advantages.

While the current trend of CMB polarization experiments has shifted to the multi-pixels, silicon wafer-based, incoherent detector approach, such as the microwave kinetic inductance detector (MKID) \citep{day03, mal10}, the conventional coherent detector approach still has its own merit, for its capability to control, detect and calibrate out the systematics.    In this regard, heterodyne polarimeters with wide bandwidths are highly desirable.   For synchrotron emitting compact sources, the interferometry array remains to be the means to reach high sensitivity, and it must adopt coherent detectors for polarization measurements.  Similar considerations apply to line emission and dust emission from compact molecular cloud cores.   All these make the conventional waveguide polarimeter device a time-honored instrument that will always be needed in future forefront millimeter/sub-millimeter telescopes.

Polarization measurements require separating the incoming radiation into two orthogonal components for the determination of Stokes parameters. Traditionally two competing devices are available for the separation of polarization, the septum polarizer and the ortho-mode transducer (OMT). An ideal septum polarizer can convert an input linear polarization wave into two circular polarization waves of equal power at two output ports. Interestingly, in a specific arrangement when the electric field of the input wave is perpendicular or parallel to the symmetric axis of the septum polarizer, the two output electric fields will be either in phase or 180 degrees out of phase, with the latter being delayed relative to the former by 90 degrees over a finite frequency interval.  It then follows that when the input is a circularly polarized wave, it will exit entirely through only one output port and the other port has a null output.  This novel feature of the septum polarizer makes it distinct from the other simpler device, OMT, which separates an input linearly polarized wave into two components of the electric field parallel and perpendicular to the device symmetry axis at two output ports \citep{wol02,men03}.  Traditionally, the septum polarizer has been known to perform well only within a relatively narrow frequency range, limited by the appearance of resonances.  On the other hand, the OMT has an advantage of being able to cover a wide bandwidth, and has been installed in modern telescopes, such as Atacama Large Millimeter/sub-millimeter Array.

A further comparison for the two kinds of devices reveals that the septum polarizer is good for measurements of Stokes $Q(\equiv [\langle E^{x}E^{x*}\rangle-\langle E^{y}E^{y*}\rangle ]/2)$ and Stokes $U(\equiv [\langle E^{x}E^{y*} \rangle+ \langle E^{y}E^{x*}\rangle ]/2)$, or the linear polarization, and the OMT good for those of Stokes $U$ and Stokes $V(\equiv i[\langle E^{x}E^{y*} \rangle - \langle E^{y}E^{x*}\rangle ]/2)$, or the circular polarization, where $\langle ...\rangle$ are the time average. The reasons are as follows.
Consider linearly polarized signals and denote the outputs of the septum polarizer to be the right-hand polarization electric field, $E^R\equiv (E^x+iE^y)/\sqrt{2}$ and the left-hand polarization electric field $E^L\equiv(E^x-iE^y)/\sqrt{2}$.  By cross-correlating $E^R$ and $E^L$, we obtain Stokes $Q$ as $\langle E^R E^{L*}+E^RE^{L*}\rangle$ and Stokes $U$ as $\langle E^RE^{L*}-E^RE^{L*}\rangle$.  In practice, one normally needs to amplify the weak incoming signals, with gains $G_R$ and $G_L$, immediately following the polarizer.
The constructed Stokes parameters are in effect $\langle G_R G_L\rangle Q$ and $\langle G_R G_L\rangle U$, assuming the gains are real.  (See Section (6) for a discussion of complex gains.)  If one has an approximate knowledge of the average gains $G_R$ and $G_L$, the Stokes $Q$ and $U$ can be determined to an acceptable accuracy.  This is what can be achieved with a septum polarizer that measures $E^R$ and $E^L$ directly. On the other hand, if one adopts the OMT, the Stokes $U$ becomes $\langle G_x G_y\rangle[\langle E^x E^{y*} \rangle+\langle E^y E^{x*} \rangle]$ and the Stokes $Q$ becomes $\langle G_x^2\rangle \langle|E^x|^2\rangle-\langle G_y^2\rangle \langle|E^y|^2\rangle$.  While Stokes $U$ can be recovered in a similar manner as that with a septum polarizer, the recovery of Stokes $Q$ has a serious problem.

This is because not only the polarized signal is already mixed with the much stronger unpolarized sky, but the amplifiers also introduce substantial unpolarized noise to the signal.  Hence $|E^x|^2$ and $|E^y|^2$ contain almost the unpolarized radiation, and the recovery of weak polarized signals is reminiscent of the determination of a very small number by subtracting two big numbers from each other, for which any small error in the two big numbers will render a poorly determined small number.  The recovery of Stokes $Q$ is therefore only possible if the amplifier gains $\langle G_x^2\rangle$ and $\langle G_y^2\rangle$ can be calibrated to a high accuracy.  However, due to the presence of gain fluctuations in amplifiers, a telescope equipped with an OMT is often difficult to yield a well-determined Stokes $Q$.   On the other hand, when the polarized signal contains only Stokes $U$ and $V$, a similar argument applies, except for replacing $E^x$ by $E^R$, $E^y$ by $E^L$, and the OMT by the septum polarizer.  But, there have rarely been pure circular polarization signals in astronomical observations; hence the OMT is normally disfavored for astronomical polarization measurements and used mostly for the measurements of Stokes $I$. (Nevertheless, a sophisticated  solution for polarization measurements with the OMT has been proposed \citep{men03}.) In spite of the great advantage of septum polarizers, they are not widely used in modern telescopes, simply because the polarizer has long been regarded as a narrow-band device. Therefore, It will be a great leap forward in the polarimeter instrumentation if this major weakness of the septum polarizer can be removed.  In this paper, we report our work specifically to address this issue.

This paper is organized as follows.  We introduce prior works on the septum polarizer in Sec.(2) and highlight the novel approach of the present work.  Sec. (3) outlines our design principles.  In Sec.(4), we report the measurement results of a polarizer fabricated for test. We convert the measurement results to the mutual leakage of Stokes parameters in Sec.(5). The leading-order calibration for reducing the Stokes $I$ leakage to other Stokes parameters is described in Sec.(6).  The conclusion is given in Sec.(7).

\section{Septum Polarizer}

The schematic of a septum polarizer is shown in Figure (1), where a stepped metal septum cuts through a circular waveguide at the midplane into two half-circular output ports. For an input $E^y$, the electric field must rotate 90 degrees to reach the output ports, and the two electric fields at the two output ports are 180 degrees out of phase (Fig. 2a). By contrast, the input $E^x$ remains intact in orientation and the two output electric fields are of the same phase (Fig. 2b). The stepped septum serves as an impedance transformer for the input $E^y$, slowing the phase velocity to create a delay relative to the input $E^x$.  To preserve the input and output powers when both components of the input electric field are present, it follows that the relative phase delay between the two components at the output ports must be $\pm$90 degrees.

While the principle of a septum polarizer has already been known for decades, the art is for the septum polarizer to cover as wide a bandwidth as possible. The very first concept of the septum polarizer is given in \citet{reg48}.  The author conceived a simple picture.  While $E^x$ propagates into the polarizer unimpeded, $E^y$, propagating along a sloping septum in a circular waveguide, must have a slower phase velocity with its orientation turning $\pm 90$ degrees on either side of the septum.  The more illuminating understanding for the functionality of the septum than previously came to light some years later. Here, the septum was regarded as a common wall \citep{che73} that had a spatially varying slot with varying cutoff frequencies \citep{sch82,beh91}.  The in-phase fields fed into the two half-waveguides cause the current to circulate in opposite directions on either side of the common wall, thereby closing the current circuit at the slot with little disturbance.  By contrast, the out-of-phase fields cause the current to flow in the same direction on the two sides of the common wall, and the slot interrupts the current, thereby altering the impedance.  In modern stepped septum polarizers, each step in the septum can actually be regarded as an individual slot.  If one alters the slot shape, the septum polarizer may be made equivalent to a single-ridge waveguide.  The ridge waveguide has been extensively studied for bandwidth enhancement and for better impedance match \citep{mon71,pat82,vah83,tuc86,bor87,bor90,bor99}.  The slower phase velocity of input $E^y$ than input $E^x$ can be understood to be due to the ridge effect, which lowers the cutoff frequency.  A ridge with a spatially varying height yields varying cutoff frequencies and controls various degrees of delay over some frequency interval.  A careful septum design can often yield 90-degree delay in $E^y$ relative to $E^x$ over some finite frequency interval.

Early developments of the septum polarizer are for phase array applications \citep{par66}.  A five-element phase array with receivers installed with sloping septum polarizers was soon after reported in a conference and the polarizer achieves $15\%$ bandwidth \citep{dav67}.  Several years later, the first journal paper using a stepped septum in a square waveguide reported that the polarizer could achieve an even wider ($20\%$) bandwidth \citep{che73}.  In the same paper, the authors claimed that the $20\%$ bandwidth was close to the bandwidth limit.  Square waveguide polarizers had been popular afterwards till 1991 when the first circular waveguide polarizer with a stepped septum was made \citep{beh91}.  Circular waveguide polarizers have the advantage that the interface to the front-end feed horn is natural without the need of a transformer, and have since been widely used in antenna arrays \citep{kum09,fra11,gal12}.

Investigations on the septum polarizer in the early days were limited to trial-and-error methods. The first analysis for a slot septum was conducted based on the Wiener-Hopf Method \citep{alb83}. Subsequently, mode-matching method, generalized transverse resonance method and finite-element method were suggested for design improvements \citep{ege85,beh91,est92}. In these studies, the excitations of high-order $TE$ and $TM$ modes presented the major challenges, thus making it necessary to adopt the single-value decomposition method to isolate the excitation modes from the fundamental mode \citep{lab92}.  It was not until 1995 that an optimized square waveguide, 4-step septum polarizer was reported \citep{bor95}.  That work provided detailed analyses for the dimensions of the steps and the thickness of the septum, and reclaimed the maximum bandwidth also to be about $20\%$.

We sum up this brief review by listing three key issues often discussed in the literature for septum polarizers. First, the bandwidth of the septum polarizer is limited primarily by that of the square (circular) waveguide. \citet{che73} already suggested that a square (circular) waveguide has a ratio of $1:1.4$ $(1:1.3)$ for the cutoff frequencies of the $TE_{01}$ $(TE_{11})$ mode and the $TE_{11}$ $(TM_{01})$ mode, therefore making the maximum bandwidth of the fundamental mode at most $34\%$ $(26\%)$.  If one further avoids the intervals of the lowest $12\%$ and the highest $2\%$ bandwidths, where the polarizer performance is difficult to control, one is left with only $20\%$ $(12\%)$ bandwidth for use.  Even with further refinements for the septum design, one can at most achieve $25\%$ $(15\%)$ bandwidth.  Exceeding this limit are the excitation of high-order modes, which can alter the phases of the fundamental modes and produce resonances.  An increase of the step number will not help, as the bandwidth limit considered above has been so fundamental that it can not be broken \citep{bor95}.

Second, impedance mismatch between the septum polarizer and the connecting waveguides can also be an issue.  There is a tendency for the septum polarizer to lower the pass band, and hence the polarizer must be made smaller by as much as $25\%$ than normal to compensate for the lowered pass band.  It thus creates an impedance mismatch problem for the polarizer's interfaces to other typical waveguides, an issue that was noted in early days \citep{par66}.

Third, the 90-degree phase shift of $E^y$ relative to $E^x$ over a wide bandwidth may be achieved either by an auxiliary Teflon thin plate next to the metal septum \citep{che73}, or by the adoption of corrugated walls in the waveguide \citep{ihm93}. The corrugated wall has long been regarded as a phase shifter \citep{sim55}.  However, these improvements are impractical in high-frequency applications. Typical dimensions of the polarizer are too small for precision arrangements of extra components and for fabrication of a complicated waveguide interior.

In this work, we report a novel design of the septum polarizer that yields good solutions to all above three issues.  We find it possible to break the aforementioned bandwidth limit by carefully optimizing the septum steps for suppressing high-order excitations and resonances over an unprecedentedly wide bandwidth.  The circular waveguide septum polarizer reported here can reach a $42\%$ bandwidth.  Moreover, the dimension of the polarizer input/out ports are only $10\%$ smaller than normal, making it easy to join other waveguide components with unsophisticated impedance transformers.  Most importantly, this septum polarizer, without any Teflon plate or corrugated wall for additional fine tuning, has been designed for high-frequency applications, specifically for W-band, and been fabricated by conventional machining tools.  An extension of this $42\%$ bandwidth design to the G-band is also envisioned, challenging the $10\%$-bandwidth G-band polarizer reported recently \citep{lea13}.

\section{Septum Design}


A circular waveguide is adopted for this polarizer.  We challenge the conventional notion of the forbidden band of the $E^y$ input, where high-order $TM_{01}$ modes are to be resonantly excited. In fact, a virtual $TM$ mode is always excited in the polarizer (Fig.(2a)) due to the intrusion of the septum, so that $E^y$ can rotate 90 degrees and be transmitted as $E^x$ at the output. However, the virtual $TM$ mode may become a real $TM_{01}$ mode and get reflected back to the input port. One of our septum design guidelines is, therefore, to prevent the virtual $TM$ mode from becoming a real $TM_{01}$ mode.  This guideline can be relatively easy to meet below the $TM_{01}$ excitation frequency.  As there is no mode conversion between the input and the reflection, the usual $\lambda/4$ rule applies.  However, the guideline becomes questionable beyond the $TM_{01}$ excitation frequency, for which the virtual $TM_{01}$ mode can become a real $TM_{01}$ mode upon reflection, a process that involves mode conversion. The difficulty in handling mode conversion has discouraged prior works from attacking the high-frequency regime, thereby drawing a bandwidth limit.

Nonetheless, careful design of the septum can still suppress the real $TM_{01}$ mode beyond the excitation frequency. While a very low level of real $TM_{01}$ mode is unavoidably excited, it is imperative that the reflected real $TM_{01}$ mode be smoothly transmitted through the input port and radiate away without a second reflection back to the polarizer interior.   This consideration leads to what we call the problem of multiple reflections. When the virtual $TM$ mode is excited, it propagates at a slower speed than the original $TE_{11}$ mode so that the input $E^y$ is delayed relative to the input $E^x$.  To create a delay exactly 90 degrees, wave internal reflections along the 5-step septum must occur for the adjustment of the tempo. Too many or too few multiple reflections in the virtual $TM$ mode can yield delays different from 90 degrees and forbid the wave from reaching the output ports.  In most frequencies, one can tune the septum steps to yield the delay close to 90 degrees.  However at some discrete frequencies, the septum tuning may fail and the virtual $TM$ modes get severely multiple-reflected (or trapped) within the polarizer to become real cavity modes, which produce resonances.  Hence, complete elimination of these cavity modes, especially the one near the $TM_{01}$ excitation frequency, is the main challenge for an extremely wide-band polarizer.

Our primary design principle for the septum is, therefore, to strictly prohibit cavity modes from occurring at the expense of allowing for some low level of real $TM_{01}$ excitations, which are to be radiated away through the input end. We begin with a size free design; when the widest percentage bandwidth is identified, we then fix the device size in accordance with the range of frequency desired. In the present case, the range is 80-116 GHz.  Given the circular waveguide, we optimize the 5-step septum with the height and the width of each step as optimization parameters.  The optimization procedure begins with the $\lambda/4$ rule for modes of different frequencies.
 To be specific, the lowest step aims to minimize the reflection of the 110 GHz fundamental, the second and third steps combined to minimize the reflection of the 85 GHz fundamental and the excitation of the 110 GHz high-order mode, the fourth step to control the 95 GHz high-order excitation and the highest step for overall performance tuning.  The above setup is used as the initial configuration for the search of optimal parameters.

Ansoft HFSS 13.0 (High Frequency Structure Simulator) was used to compute the scattering (S) parameters that serve as the optimization indicators. The left-right symmetry of the polarizer allows us to compute only half of the space so as to speed up the parameter search.  After the optimal parameters are found, we employ the full space simulation to re-compute S parameters as a confirmation check, and to determine the precision tolerances of fabrication. Note that if high-order modes were incorrectly not allowed to be present as input/output eigen-modes in the HFSS simulation, these high-order excitations would not be able to transmit away from the polarizer, producing a large number of multiple reflections, and the results would always yield erroneous strong resonances.

We indeed find an optimal solution free of resonance, not only across the entire W-band but also extending beyond 120 GHz.  The optimized polarizer aperture is found to be 2.5 mm with the $TM_{01}$ excitation frequency at 93 GHz.  The optimized septum width is found to be $8\%$ of the polarizer aperture, 0.2 mm, for good performance at high frequency and for the mechanical rigidity of a bronze septum.  The optimized HFSS results are shown in Fig. (3), where the $TE_{11}$ modes $(E^x,E^y)$ at the common port are denoted as mode 1, the $TE_{11}$ mode at $R/L$ ports as mode 2, the excited $TM_{01}$ mode as mode 3, and the excited $TE_{21}$ mode as mode 4.  Here, the $S$ parameter is defined as:
\\
\begin{equation}\label{1}
  S_{ij}(E^{x,y})=10\log_{10}\frac{\left|E^{x,y}(mode_{i})\right|^2}{\left|E^{x,y}(mode_{j})\right|^2}_{.}
\end{equation}
\\
It is found from Fig. (3) that the input reflections $S_{11}(E^x)$ and $S_{11}(E^y)$
are under -20 dB and $S_{31}(E^y)$ under -14 dB within 94-118 GHz. The high-order mode $TM_{01}$ cannot be excited by the $E^x$ input and indeed $S_{31}(E^x)$ is close to zero.  The transmission $S_{21}(E^x)$ is nearly perfect;
however, $S_{21}(E^y)$ has some loss due to energy conversion to the high-order mode beyond 93 GHz, and the loss is at most 0.2 dB in between 94-100 GHz and 0.3 dB at 118 GHz.
We find that the excitation of $TM_{01}$ mode is unavoidable beyond its cutoff frequency, and beyond 118 GHz an even higher-order mode $TE_{21}$ begins to be significantly excited for both $E^y$ and $E^x$ inputs.  Nevertheless one can manage to keep the high-order excitation level low at least up to 118 GHz.  In particular, the simulation results in Fig. (3) show that avoiding resonances is achievable over a very wide frequency range (75 - \textgreater120 GHz), thus opening a new regime of operation for a waveguide septum polarizer.

To provide an evidence for the septum design to be close to the optimum, we change the waveguide diameter by $\pm8\%$, with the septum shape and dimension to remain intact. Figure (4) depicts the HFSS simulation results of these changes for the $E^{y}$ inputs that are more sensitive to optimization parameters than the $E^{x}$ inputs. The obvious differences are the shifted frequencies that scale inversely with the diameter. We rescale the frequency axis of $S_{21}$ in Fig. (3) by $\pm8\%$ and plot them in Fig. (4a and 4b) for detailed comparisons of the altered configurations with the optimal configuration. The change in diameter yields slightly declining performance, demonstrating that the present polarizer design is very close to the optimum.

\section{Measurements}

Before proceeding to the presentation of measurement results, we find it important to stress the arrangements before the signal enters the polarizer from the common port. This polarizer operates in the frequency range beyond the cutoff of $TM_{01}$ excitation.  Despite that we can manage to suppress the high-order excitation to a great degree, there is still some low level of $TM_{01}$ mode that gets reflected back to the front-end devices.  If the front-end devices do not allow the $TM_{01}$ mode to radiate, the reflection of it from the front-end back to the polarizer interior will lead to a cavity effect and creates spurious new resonances. It is therefore essential that the front-end device permits total transmission of $TM_{01}$ modes. The condition is naturally fulfilled in the telescope setting since the polarizer is preceded by a feed horn, which allows the reflected $TM_{01}$ mode to radiate away.  However, the radiation boundary condition cannot be satisfied when we perform measurements by connecting the common port of the polarizer to the rectangular waveguide of the measurement device.  In this case, the excitation mode is totally reflected back to the polarizer, thereby producing new resonances.

The polarizer is measured by the HP 8510 vector network analyzer (VNA) that covers 75-115 GHz.  Two types of measurements are made.  For measurement A, the common port is the output port of the polarizer, which is connected to a Potter feed horn \citep{lee12}, and the R and L ports are connected to the two ports of VNA.  For measurement B, one port of VNA is connected to the common port via a rectangular-to-circular waveguide transition adapter and the other port of VNA to the R (L) port, with the L (R) port properly terminated.  Measurement A tests the performance of the R/L port return loss and mutual isolation.  If a high-order mode that is excited away, and there can be no telling of the high-order excitation in measurement A.  However, if there are internal multiple reflections, i.e., cavity modes, inside the polarizer, measurement A can reveal the resonances.   On the other hand, measurement B must use a transition adapter in between the VNA and the common port, and can cause a serious problem in reflecting the excited $TM_{01}$ mode back to the polarizer, creating the otherwise absent resonances in the VNA measurement.  Nevertheless, if one ignores the responses at some discrete resonances and reads only the continuum results, measurement B can provide the full characteristics, thus full Stokes parameters, of the receiver polarizer.

(a)	Measurement A:

Figure (5) summarizes the results.  First of all, no resonance appears in this measurement. In the interval between 85 GHz and 115 GHz, the return loss and the isolation are largely below -20 dB. The slight rise of return loss to -17 dB in the low-frequency interval 77-85 GHz is due to the slight mismatch between the WR10 rectangular waveguide of the VNA and the polarizer semi-circular waveguide. We verify the measurement results by simulating the exact measurement configuration with HFSS, finding good agreements especially in low frequency where the waveguide mismatch prevails. This is a minor problem that can be easily corrected by a transition adapter.

In the past, the polarizer was found to show resonances above the excitation frequency of high-order modes in measurement A, from which the bandwidth limit mentioned in Sec. (2) was set.  For the detection of resonances, A is a reasonable measurement because when multiple reflections of high-order modes in a cavity occur, part of the high-order modes should be converted to the fundamental mode back to the input and isolation ports.   It can thus reveal the presence of resonances, though the conversion efficiency may be low.  We find no resonance to be present in measurement A, providing an evidence against any cavity mode inside the septum polarizer.  Full justification of this claim requires the input to be injected from the common port, and it leads us to measurement B.

(b)	Measurement B:

As mentioned earlier, measurement B cannot be free of problems since the VNA rectangular waveguide port produces multiple reflections of $TM_{01}$ modes in the polarizer.  Hence spurious resonances are always present in this measurement.  To circumvent this difficulty, our solution is to change the length of the rectangular-to-circular transition waveguide and examine whether any identical resonance exists regardless of such a change.  The rationale behind this approach is that if any internal cavity mode is to exist, its resonance frequency should be independent of the length changes of the external reflector.  Our measured results will be further checked against the HFSS simulations to ensure the correctness of the interpretation.

Figure (6) presents results for the $E^x$ input and for the $E^y$ input, respectively.  Here the length of the circular waveguide section in the transition adapter is chosen to be the shortest possible, 0.2 mm. The result for the $E_y$ input reveals three resonances at 95.2 GHz, 104.2 GHz and 110.2 GHz, and the HFSS simulation exactly yields the same resonances.  The result for $E_x$ reveals an unexpected single resonance at 95 GHz.  This odd resonance is actually produced from the mutual leakage between $E^x$ and $E^y$, due to axis misalignment by 1.8 degrees at the interface between the rectangular-to-circular transition adapter and the polarizer.  The misalignment has also been verified by the HFSS simulation, shown in Fig. (6) as well.  We also note in Fig. (6) that the power imbalance of $R$ and $L$ ports in input $E^x$ and in input $E^y$ measurements tends to be opposite.  This result is also caused by the 1.8-degree axis misalignment, a problem that can be further verified by the mutual leakage of Stokes $Q$ and $U$ discussed in the next section.

Other than these discrete resonances, the polarizer performs well in the continuum, with about 0.2-0.3 dB additional power loss compared to the ideal polarizer simulation results. This additional loss is partly caused by the transition adapter preceding and a splitter following the polarizer.  The output power imbalance between R and L ports in the continuum is also small, within 0.2 dB on average and 0.3 dB maximum, despite that half of the power imbalance is produced by irrelevant axis misalignment.  (In real telescope applications, the rectangular-to-circular transition adapter will not be present, and the axis misalignment will not be an issue of concern.)
The phase at every frequency has also been measured but is not shown here.  These phase measurements are important for our further examination of the mutual leakage among four Stokes parameters in the next section.

Figure (7) basically presents the same measurements as Fig. (6), but with different sets of rectangular-to-circular transition adaptors of 5 mm and 10 mm in length. The resonances in this case are more closely packed than the previous case, therefore somewhat deteriorating the performance in the continuum.  Hence we shall focus on identification of the resonances $\it{per se}$.  The resonances in the measurements are located at 92.4 GHz, 95.6 GHz, 101.2 GHz, 105.4 GHz, and 109.6 GHz for the 5 mm adaptor and 92.0 GHz, 93.4 GHz, 95.8 GHz, 99.4 GHz, 103.4 GHz, 106.0 GHz, 109.6 GHz, and 114.2 GHz for the 10 mm adaptor.  Together with Fig. (6), we find that none of these resonances is in common in all three cases of different external cavity lengths, indicative of no internal cavity mode in the polarizer.  We also find that all measured resonances coincide with the simulation resonances.  The confirmation of measurements by simulations further reinforces our confidence that this polarizer is free of resonance over the measurement range
from 75 GHz to 115 GHz.  The faithful reproduction of the measured results by simulations in turn makes it believable that the polarizer design should be free of resonance even beyond 120 GHz, as indicated by the simulation presented in Fig. (3).




\section{Polarization Leakage}

A number of minor imperfections in the polarizer lead to small leaks among different Stokes parameters. In view of the weak polarized signal in the presence of the stronger unpolarized source, the primary concern of a polarizer is the leakage from Stokes I to other three Stokes parameters. Other lesser critical concerns are the mutual leakage among the three polarized components. Since the observed polarized radiation is mostly linearly polarized, the Stokes V is zero, and the polarization mutual leakage is between Stokes $Q$ and $U$. As long as the $Q-U$ leakage is controlled within the few-percent level, the performance of the polarizer is considered to be acceptable \citep{lei02}. However, in very demanding observations such as the B-mode polarization observations of the CMB radiation, it is the level of $Q-U$ leakage that sets the sensitivity limit of an instrument \citep{hu03,ode07}. Below, we compute the mutual leakage of the four Stokes parameters from the data of measurement B with the 0.2 mm transition adapter.

The output complex electric fields at the $R$ and $L$ ports for an ideal septum polarizer are expressed as:

\begin{subequations}
\begin{eqnarray}
&& E^{R} =
\frac{1}{\sqrt{2}}(E^{x}+iE^{y}) \\
&& E^{L} =
\frac{1}{\sqrt{2}}(E^{x}-iE^{y}), \\
\nonumber &&
\end{eqnarray}
\end{subequations}

\hspace{-27pt} for an ideal septum polarizer. Consider different polarizers inside a pair of receivers $(m,n)$.  The visibility is known as the time-averaged cross-correlation of the complex electric fields incident to receivers $m$ and $n$, and the correlation responses are obtained as:
\begin{subequations}
\begin{eqnarray}
  \left \langle{E^{R}_{m}E^{R*}_{n}}\right \rangle &=& \frac{1}{2}[(E^{x}_{m}E^{x*}_{n}+{E^{y}_{m}E^{y*}_{n}})-i(E^{x}_{m}E^{y*}_{n}-E^{y}_{m}E^{x*}_{n})]=\frac{1}{2}(I+V) \\
  \left \langle{E^{L}_{m}E^{L*}_{n}}\right \rangle &=& \frac{1}{2}[(E^{x}_{m}E^{x*}_{n}+{E^{y}_{m}E^{y*}_{n}})+i(E^{x}_{m}E^{y*}_{n}-E^{y}_{m}E^{x*}_{n})]=\frac{1}{2}(I-V) \\
  \left \langle{E^{R}_{m}E^{L*}_{n}}\right \rangle &=& \frac{1}{2}[(E^{x}_{m}E^{x*}_{n}-{E^{y}_{m}E^{y*}_{n}})+i(E^{x}_{m}E^{y*}_{n}+E^{y}_{m}E^{x*}_{n})]=\frac{1}{2}(Q+iU) \\
  \left \langle{E^{L}_{m}E^{R*}_{n}}\right \rangle &=& \frac{1}{2}[(E^{x}_{m}E^{x*}_{n}-{E^{y}_{m}E^{y*}_{n}})-i(E^{x}_{m}E^{y*}_{n}-E^{y}_{m}E^{x*}_{n})]=\frac{1}{2}(Q-iU) \\
\nonumber &&
\end{eqnarray}
\end{subequations}

The co-polar ($E^RE^R,E^LE^L$) correlations provide information about $I$ and $V$ and the cross-polar ($E^RE^L,E^LE^R$) correlations about $Q$ and $U$.   For simplicity we consider a radiometer polarizer as an example with $m=n=1$ and drop the receiver indices.   We model the imperfect responses of the $E^R$ and $E^L$ outputs after the amplifiers as

\begin{subequations}
\begin{eqnarray}
&&\tilde{E}^{R}=\frac{G_{R}}{\sqrt{2}}[(1-\Delta_R)E^{x}+ie^{i\alpha_R}(1-\varepsilon_R)E^{y}] \\
&&\tilde{E}^{L}=\frac{G_{L}}{\sqrt{2}}[e^{i\alpha_\Delta}(1-\Delta_L)E^{x}-ie^{i\alpha_L}(1-\varepsilon_L)E^{y}] \\
\nonumber &&
\end{eqnarray}
\end{subequations}

where $\Delta_{R,L}$ and $\varepsilon_{R,L}$ denote the magnitude losses of $E^{x}$ and $E^{y}$ at ports R and L, and $\alpha_{\Delta}$, $\alpha_{R}$ and $\alpha_{L}$ denote the phase errors in reference to $E^{x}$ at the $R$ port.   Here, $\Delta_{R,L}$, $\varepsilon_{R,L}$, $\alpha_{\Delta}$, $\alpha_{R}$ and $\alpha_{L}$ are of the same order of smallness $O(\eta)$, Here, $\eta < 2\%$ from the VNA measurements, and the leading-order corrections suffice to compute the leakage.  We also take the amplifier gains, $G_R$ and $G_L$, to be real. This is because the relative phase between the two complex gains and the relative path delay in $\tilde{E}^R$ and $\tilde{E}^L$ can be pre-determined and calibrated out. Hence after the phase calibration, the gain can be made a real quantity. The correlations of the two amplified electric fields now become:

\begin{subequations}
\begin{eqnarray}
&&\tilde Q\equiv\Re \left \langle{2\tilde{E}^{R}_{1}\tilde{E}^{L*}_{1}}\right \rangle \nonumber \\
&&= \langle\left| G_{R}G_{L} \right|\rangle\ \Big \{- \textstyle\frac{1}{2}[(\Delta_L-\varepsilon_L)+(\Delta_R-\varepsilon_R)]I\\
&&+ \{ \textstyle1-\frac{1}{2}[(\Delta_R+\Delta_L)+(\varepsilon_R+\varepsilon_L)]\}Q  \\
&&+ \textstyle\frac{1}{2}[(\alpha_L-\alpha_R)+\alpha_\Delta)]U\\
&&- \textstyle\frac{1}{2}[(\Delta_L-\varepsilon_L)-(\Delta_R-\varepsilon_R)]V \Big \}, \\
\nonumber &&
\end{eqnarray}
\end{subequations}

\begin{subequations}
\begin{eqnarray}
&&\tilde U\equiv\Im \left \langle{2\tilde{E}^{R}_{1}\tilde{E}^{L*}_{1}}\right \rangle \nonumber \\
&&= \langle\left| G_{R}G_{L} \right|\rangle\ \Big \{ -\textstyle\frac{1}{2}[(\alpha_\Delta-\alpha_L)+\alpha_R]I\\
&&- \textstyle\frac{1}{2}[(\alpha_L-\alpha_R)+\alpha_\Delta]Q \\
&&+ \{\textstyle1-\frac{1}{2}[(\Delta_R+\Delta_L)+(\varepsilon_R+\varepsilon_L)]\}U\\
&&- \textstyle\frac{1}{2}[(\alpha_\Delta-\alpha_L)-\alpha_R]V \Big \},\\
\nonumber &&
\end{eqnarray}
\end{subequations}

\begin{subequations}
\begin{eqnarray}
&&\tilde V\equiv\left \langle{\tilde{E}^{R}_{1}\tilde{E}^{R*}_{1}}\right \rangle-\left \langle{\tilde{E}^{L}_{1}\tilde{E}^{L*}_{1}}\right \rangle \nonumber \\
&&= \textstyle\frac{1}{2}\Big \{\langle\left| G_{R} \right|^{2}\rangle (1-\Delta_R-\varepsilon_R)-\langle\left| G_{L} \right|^{2}\rangle (1-\Delta_L-\varepsilon_L)\Big \}I\\
&&+\textstyle\frac{1}{2}\Big \{\langle\left| G_{L} \right|^{2}\rangle(\Delta_L-\varepsilon_L)-\langle\left| G_{R} \right|^{2}\rangle(\Delta_R-\varepsilon_R)\Big \}Q\\
&&+\textstyle\frac{1}{2} \Big \{\langle\left| G_{L} \right|^{2}\rangle(\alpha_\Delta-\alpha_L)-\langle\left| G_{R} \right|^{2}\rangle\alpha_R \Big \} U\\
&&+\textstyle\frac{1}{2}\Big \{\langle\left| G_{R} \right|^{2}\rangle(1-\Delta_R-\varepsilon_R)+\langle\left| G_{L} \right|^{2}\rangle(1-\Delta_L-\varepsilon_L) \Big \}V,\\
\nonumber &&
\end{eqnarray}
\end{subequations}

\begin{subequations}
\begin{eqnarray}
&&\tilde I\equiv \left \langle{\tilde{E}^{R}_{1}\tilde{E}^{R*}_{1}}\right \rangle+\left \langle{\tilde{E}^{L}_{1}\tilde{E}^{L*}_{1}}\right \rangle \nonumber \\
&&= \textstyle\frac{1}{2}\Big \{\langle\left| G_{R} \right|^{2}\rangle (1-\Delta_R-\varepsilon_R)+\langle\left| G_{L} \right|^{2}\rangle (1-\Delta_L-\varepsilon_L)\Big \}I\\
&&-\textstyle\frac{1}{2}\Big \{\langle\left| G_{L} \right|^{2}\rangle(\Delta_L-\varepsilon_L)+\langle\left| G_{R} \right|^{2}\rangle(\Delta_R-\varepsilon_R)\Big \}Q\\
&&-\textstyle\frac{1}{2} \Big \{\langle\left| G_{L} \right|^{2}\rangle(\alpha_\Delta-\alpha_L)+\langle\left| G_{R} \right|^{2}\rangle\alpha_R \Big \} U\\
&&+\textstyle\frac{1}{2}\Big \{\langle\left| G_{R} \right|^{2}\rangle(1-\Delta_R-\varepsilon_R)-\langle\left| G_{L} \right|^{2}\rangle(1-\Delta_L-\varepsilon_L) \Big \}V.\\
\nonumber &&
\end{eqnarray}
\end{subequations}

The above four expressions contain the leading order Stokes parameters, $Q$, $U$, $V$ and $I$, followed by the leakage from other three Stokes parameters on the order $O(\eta)$. The leakage obeys a symmetry principle, as a result of the scattering matrix being unitary or the quantity $I^2-(Q^2+U^2+V^2)$ being an invariant, if no loss were to occur. The leakage coefficients between $Q$ and $U$, Eq. (5.4c) and Eq. (5.5b), are the same in magnitude but opposite in sign, representing the antenna principal axes are not perfectly aligned with the polarizer axes and rotate by a small amount.   The coefficients of leakage from $I$ to $Q$, $U$ and $V$, Eqs. (5.4a), (5.5a) and (5.6a), respectively, are the same as those of the leakage from
$Q$, $U$ and $V$ to $I$, Eqs. (5.7b), (5.7c) and (5.7d), when the two gains are the same $G_R=G_L$.  The leakage between $Q$ and $V$ and that between $U$ and $V$ also have the same magnitudes but opposite signs when $G_R=G_L$.   Finally, the net loss in Stokes $I$, Eq. (5.7a), is the same as the loss in $Q$, Eq. (5.4b), in $U$, Eq. (5.5c), and in $V$, Eq. (5.6d), again when $G_R=G_L$.  Hence without amplifiers, the leakage in a septum polarizer is determined by 6 parameters, 3 from the amplitude imbalance, i.e., $\Delta_R+\Delta_L-\epsilon_R-\epsilon_L$, $\Delta_R-\Delta_L+\epsilon_R-\epsilon_L$, and $\Delta_R-\Delta_L-\epsilon_R+\epsilon_L$, and 3 from the phase imbalance, i.e., $\alpha_\Delta-\alpha_R-\alpha_L$, $\alpha_\Delta+\alpha_R-\alpha_L$, and $\alpha_\Delta-\alpha_R+\alpha_L$.   If the polarizer is lossy, there will be an additional
parameter, $\Delta_R+\Delta_L+\epsilon_R+\epsilon_L$, that gives uniform suppression of all 4 Stokes parameters.

Take the VNA measurement data for two individual septum polarizers fabricated with the same design, and we can compute various leakage coefficients according to the formula given above.  As the measurements involve only VNA with no amplifiers, we let $G_R=G_L=1$.  In Figs. (8)\&(9), we plot the leakage from $I$ to other three polarization components and the polarization mutual leakage.  The measurement results are consistent for the two polarizers, and they are summarized in Table I.
Clearly the leakage are systematically different below and above the excitation frequency at 95 GHz, despite they are both small.  Given the finite line widths of the three resonances present in Fig. (6), the measured results of the nearby continuum are likely contaminated by the poor responses at the resonances.  Therefore, the leakage should be regarded as pessimistic, and the actual performance should be better than the results indicated here.  We additionally find that the mechanical requirement of the polarizer has no major bottleneck, judging from the performance consistency of the two polarizers.

\section{Calibration for Removing Leakage from Stokes I}

Given the estimated leakage of this polarizer from the large Stokes I to other three small Stokes components, one can further perform calibration at the system level to further reduce the $I$ leakage of $2\%$ level.  To the leading order, only the leakage from the much stronger unpolarized component to the polarized component is to be calibrated out. Higher-order calibrations are possible, but this subject is quite involved and will not be discussed here.  Again we take the simple case of a radiometer, where the single receiver outputs, $G_RE^R$ and $G_LE^L$, are to be correlated to obtain the four Stokes parameters. In contrast to Sections (1)\&(5), here $G_R$ and $G_L$ are now taken to be complex, including system phase delays.   With an unpolarized source as an input to a perfect receiver, one would obtain a finite value for Stokes I and zero values for other Stokes components.  Non-zero values in other Stokes components in a real system represent the instrument leakage that is to be removed.   The first part of system calibration can be conducted in the laboratory and and determine the coefficients present in Eq. (5.4), (5.5) and (5.6) from the four non-zero Stokes parameters.  The second part of calibration is to be conducted in the field so that the Stokes $I$ leakage can be subtracted away from the observed polarized components.

The magnitudes of the gains $G_R$ and $G_L$ can be determined from the power measurements of the two outputs $|G_RE^R|^2$ and $|G_LE^L|^2$, with the receiver exposed to an unpolarized source of known temperature. Determination of the relative phase between $G_R$ and $G_L$ is tricky, requiring a variable delay to alter the relative phase between $G_RE^R$ and $G_LE^L$. For a finite bandwidth source, the delay cross correlation between $G_RE^R$ and $G_LE^L$ will create fringes with an amplitude modulation as a function of the delay. The zero delay between $G_R$ and $G_L$ is one for which both amplitude modulation and fringe are the maximum. When calibration is to be performed with frequency resolution, digitization of data is required for the determination of complex gains per frequency, $G_R(\nu)$ and $G_L(\nu)$.  All the above tasks can be conducted in a well-controlled laboratory and hence the complex gain $G_R$ and $G_L$ can be measured at high precision.

Once the complex gains are determined, the leading order leakage of Stokes I to other three Stokes components, Eqs. (5.4a), (5.5a) and (5.6a), can then be measured with an unpolarized source.  Unlike the VNA measurement results presented in Sec.(5), which are affected by spurious resonances, the quantities $(\Delta_R+\Delta_L)-(\epsilon_R+\epsilon_L)$, $\alpha_\Delta+\alpha_R-\alpha_L$ and $(\Delta_R+\epsilon_R)-(\Delta_L+\epsilon_L)$ can be directly determined at the system level.

In the field, where other three Stokes parameters are much smaller than Stokes I by a factor $O(\delta)$, with $\delta << 1$, it is necessary to measure Stokes $I$ in order to construct the $I$ leakage according to Eqs. (5.4a), (5.5a) and (5.6a).   However, the removal of $I$ leakage to Stokes $V$ may be difficult to perform in the field due to the gain fluctuations, and therefore a poorly determined Stokes $V$ is expected.  After the calibration, the leakage from Stokes $I$ to Stokes $Q$ and $U$ can theoretically be removed up to $O(\eta^2)$ compared to the polarized components, since $I$ can in turn be contaminated by leakage from $Q$, $U$ and $V$ (c.f. Eq. (5.7)).  In practice, the accuracy of the measured Stokes $I$ places a limit on the degree of removal of the $I$ leakage.

For example, assume that the gains can ideally be determined accurately after a long integration, and the weak linear polarization signals are detected with a signal-to-noise ratio, S/N$=10$.  The Stokes $I$ must have already been measured with a high S/N about $O(10\delta^{-1})$.  With a measurement error of order $\delta/10$ in Stokes $I$, the leakage to polarized components from the unpolarized component can be removed up to $O(\eta/10)$ of the polarized components.  Despite that the residual is not as small as $O(\eta^2)$, it is still much smaller than the polarization mutual leakage $O(\eta)$.
On the other hand, the mutual leakage among Stokes $Q$ and $U$ cannot be reduced by the leading-order calibration since $Q$ and $U$ are weak with a relatively low $S/N$, unlike the strong $I$.  It is therefore up to the performance of the polarizer to limit the leakage, and in our case the $Q-U$ mutual leakage is $<\pm 1\%$.

In sum, after the leading-order removal of the Stoke $I$ leakage, the polarized components can be made accurate up to order $O(\eta)$ contributed primarily by the $Q-U$ mutual leakage.  Higher-order calibration is possible with an even deeper integration, provided that the receiver system is sufficiently stable.

\section{Conclusion}

In this paper, we report a novel design of the septum polarizer that has a $42\%$ well-performed bandwidth, from 77 GHz to 118 GHz, as opposed to the conventional notion of about $20\%$ maximum bandwidth. The conventionally alleged maximum bandwidth was derived primarily from a consideration of a limit set by the cutoff frequency of the fundamental modes and the excitation frequency of high-order modes. Our polarizer, adopting a circular waveguide and housing a 5-step septum, is able to break the upper bandwidth limit and extends the usable frequency into the range where high-order $TM_{01}$ modes are excited.  This is made possible because we have uncovered an under-explored regime in which $TM_{01}$ excitations can be severely suppressed, and $TM_{01}$ resonances be entirely eliminated.   Particularly near the $TM_{01}$ excitation frequency, the septum can manage to avoid multiple reflections of the long wavelength modes inside the polarizer. In addition, the mutual leakage among all four Stokes parameters has been measured. It shows that this septum polarizer performs well, having $I$ to $Q$, $U$ leakage less than $2\%$ and $Q-U$ mutual leakage less than $\pm 1\%$ in almost all frequencies.

A few dozen of polarizers of our design have been fabricated by conventional precision machining with $\pm 5$ $\mu m$ tolerance, and most of them have similar performances as the two modules reported here.  Due to the simplicity of the polarizer without any complicated component to assist widening the bandwidth, this polarizer sets a milestone for the instrumentation of polarization measurements.

This septum polarizers will be installed in an upgraded National Taiwan University (NTU)-Array prototype, where each receiver is equipped with 19 pixels of coherent detectors.  The $80-116$ GHz signals collected by the upgraded NTU-Array are to be processed by digital correlators that are designed to cover simultaneously the 36 GHz bandwidth for all pixels in all receivers with frequency resolution down to 100 kHz using software correlation.  In the context of this work, the backend digital processing capability of the telescope can help calibrate these polarizers with fine frequency resolution.




\acknowledgements{
We thank Dr. D.C. Niu and Dr.  C.C. Chiong for their kind assistance on our VNA measurements and ASIAA for granting us an access to their HP8510.   Valuable discussions with Dr. K.Y. Lin
are also acknowledged. This project is supported in part by the NSC grants, 100-2627-E-002-002 and 100-2112-M-002-018.}

\clearpage



\begin{figure}
\hspace{100pt} \includegraphics[angle=90,scale=.60]{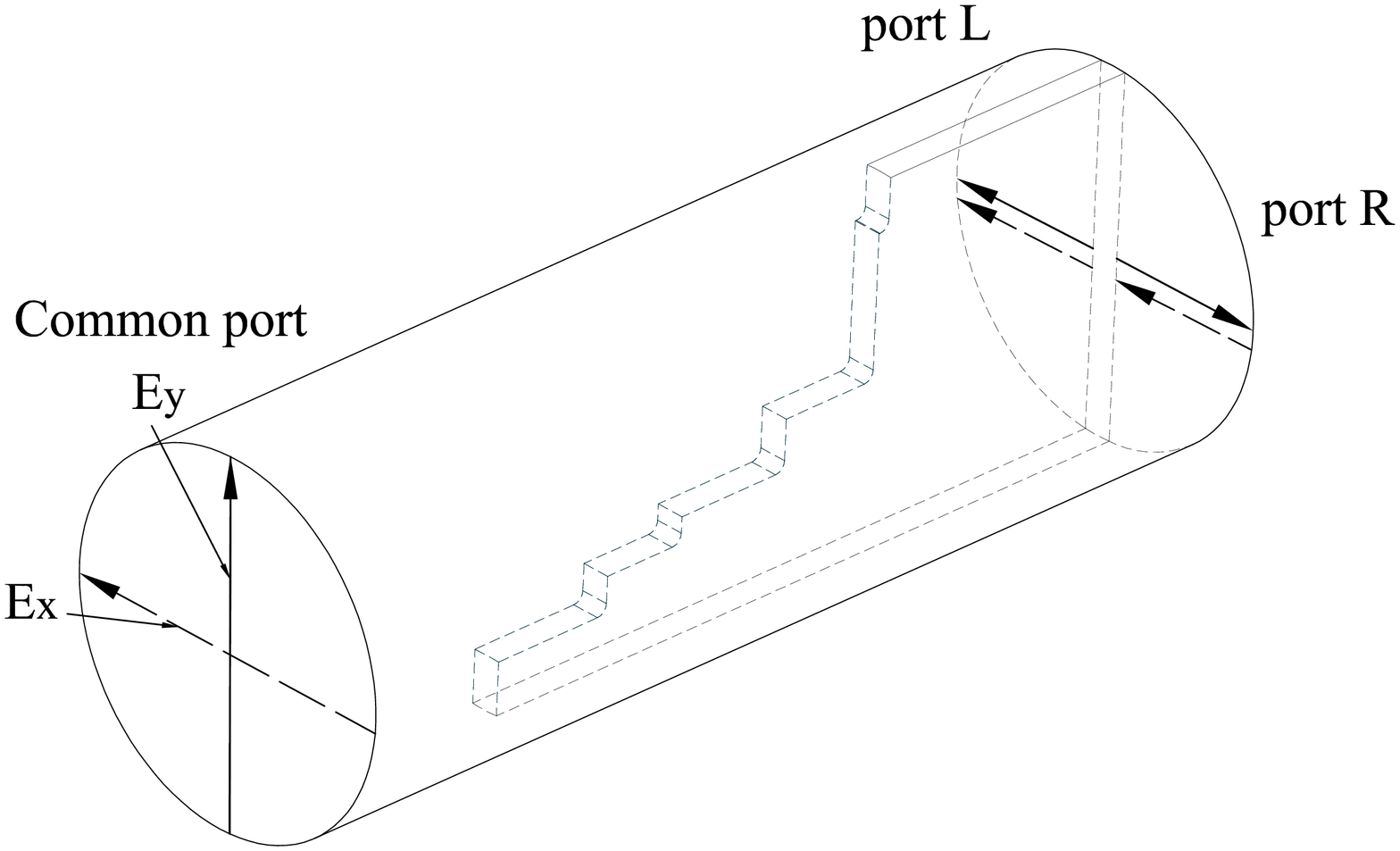}
\vspace{50pt}
\caption{A typical configuration of the stepped septum polarizer in a circular waveguide. Vertical component ($E^y$) and horizontal component ($E^x$) of the electric field fed into the common port are separated by the septum to become the right-hand polarization component output at port R and the left-hand polarization component output at port L. \label{fig1}}
\end{figure}

\begin{figure}
\vspace{30pt}
\hspace{150pt} \includegraphics[angle=90,scale=.40]{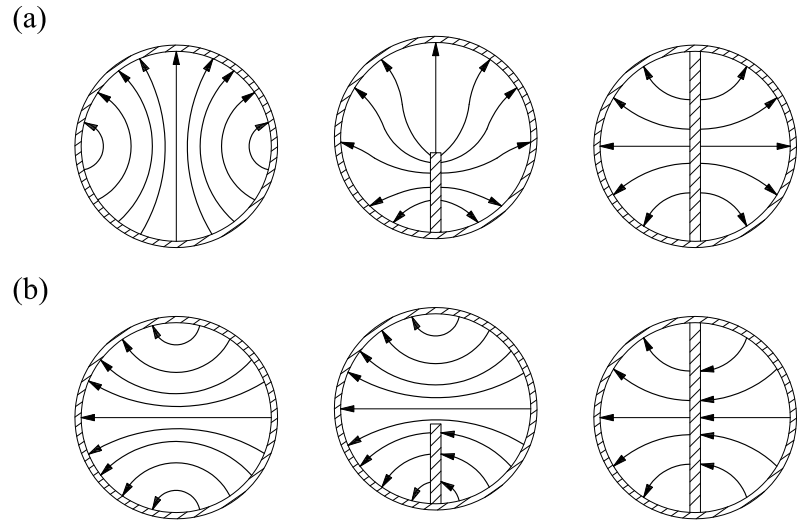} \\
\\
\\
\vspace{10pt}
\caption{Field distributions of an $E^y$ input (a) and an $E^x$ input (b).  The electric current smoothly circulates in opposite directions on either side of the common wall (septum) for the $E^x$ input as if no septum ever existed.  But the current flows in the same direction on either side of the common wall for the $E^y$ input, so that the septum top edge becomes a stagnation point for charge accumulation. Therefore, a virtual $TM_{01}$ mode, which has primarily the radial electric field, is excited.  A good septum is able to re-convert the virtual $TM_{01}$ mode back to the $TE_{11}$ mode on its exit to output ports. \label{fig2}}

\end{figure}

\begin{figure}
\includegraphics[angle=270,scale=.60]{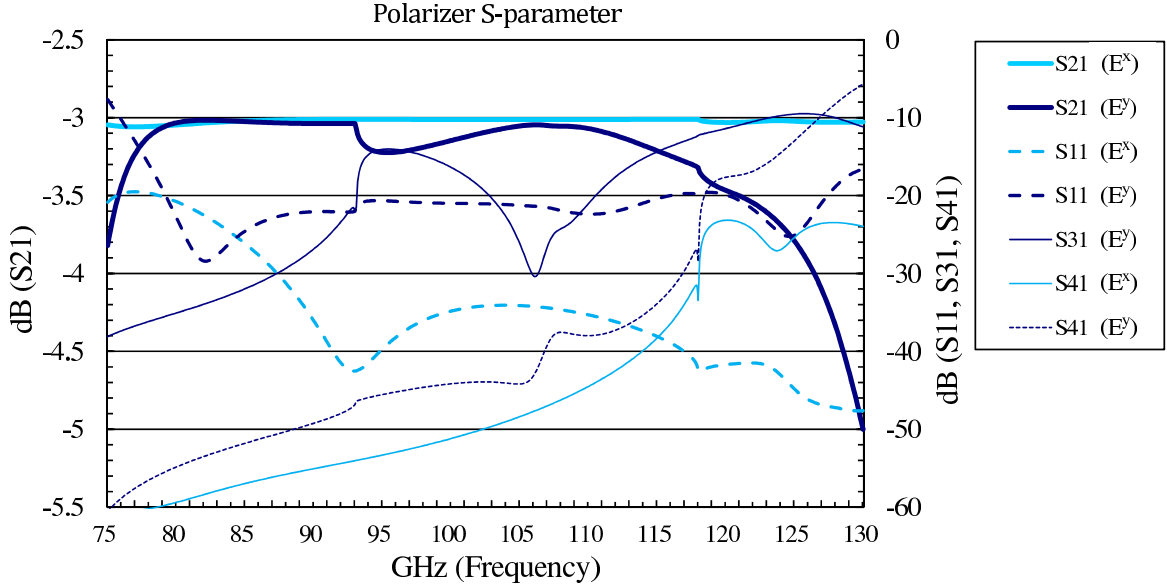}
\vspace{30pt}
\caption{Simulation results of the transmission $S_{21}$ and the reflection $S_{11}$ of the fundamental $TE_{11}$ mode,
and of the reflected $TM_{01}$ ($S_{31}$) and $TE_{21}$ ($S_{41}$) modes, for our optimized septum polarizer. These simulation results are for an ideal polarizer, where the left-right symmetry is obeyed.  For the $E^y$ input, the $TM_{01}$ is seen to be well suppressed except near the $TM_{01}$ cutoff frequency 93 GHz.  But even near this frequency the suppression is still good at the -14 dB level with an insertion loss 0.2 dB.  Most impressively, this polarizer design has been tuned to eliminate all resonances across the entire W-band and beyond $120$ GHz. The even higher-order modes $TE_{21}$ begins to be excited beyond 118 GHz for both $E^y$ and $E^x$ inputs, and the transmission $S_{21}$ for the $E^y$ input deteriorates rapidly beyond 120 GHz.
\label{fig3}}
\end{figure}

\begin{figure}
\vspace{50pt}
\includegraphics[angle=270,scale=.50]{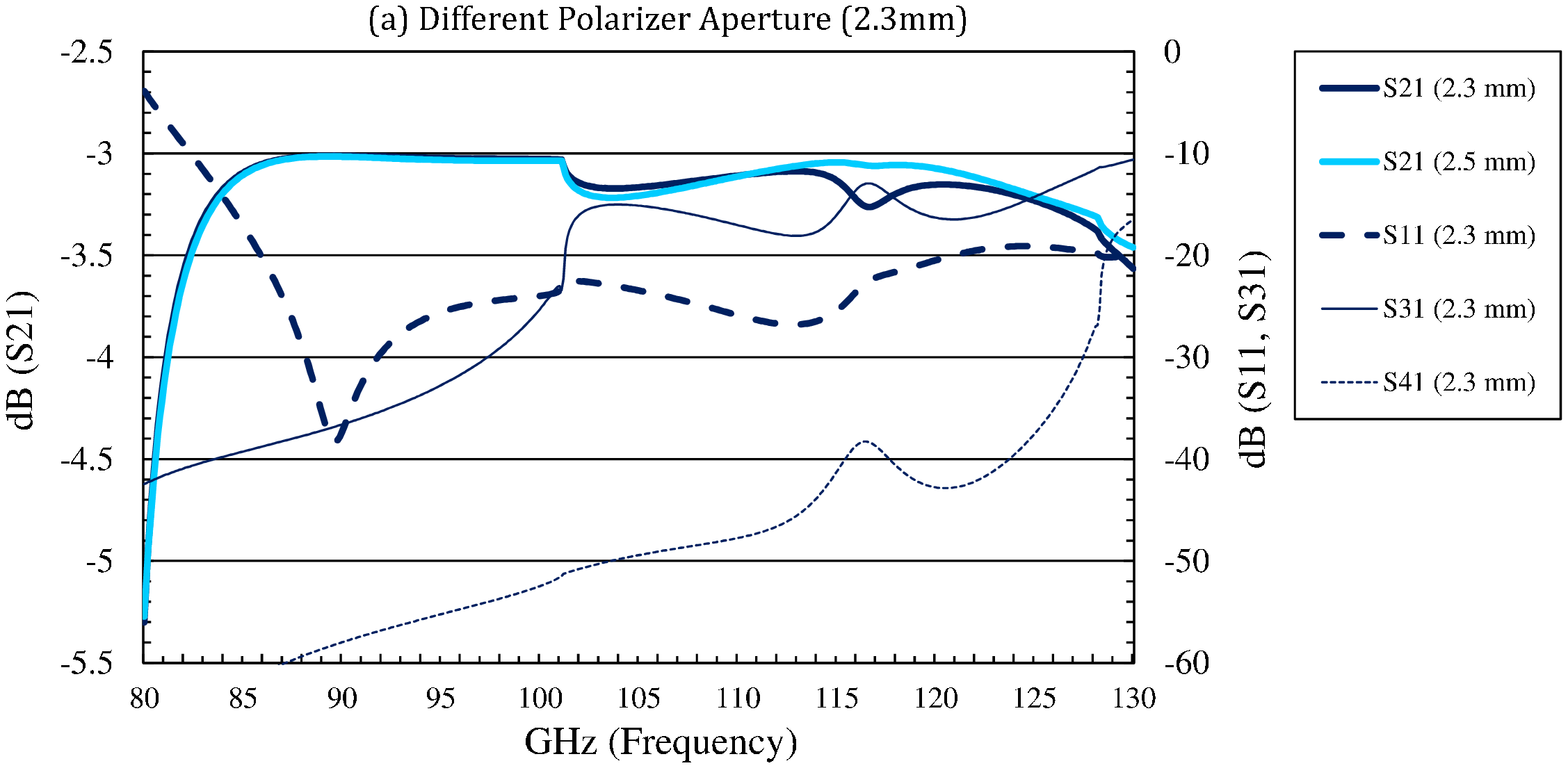} \\
\\
\includegraphics[angle=270,scale=.51]{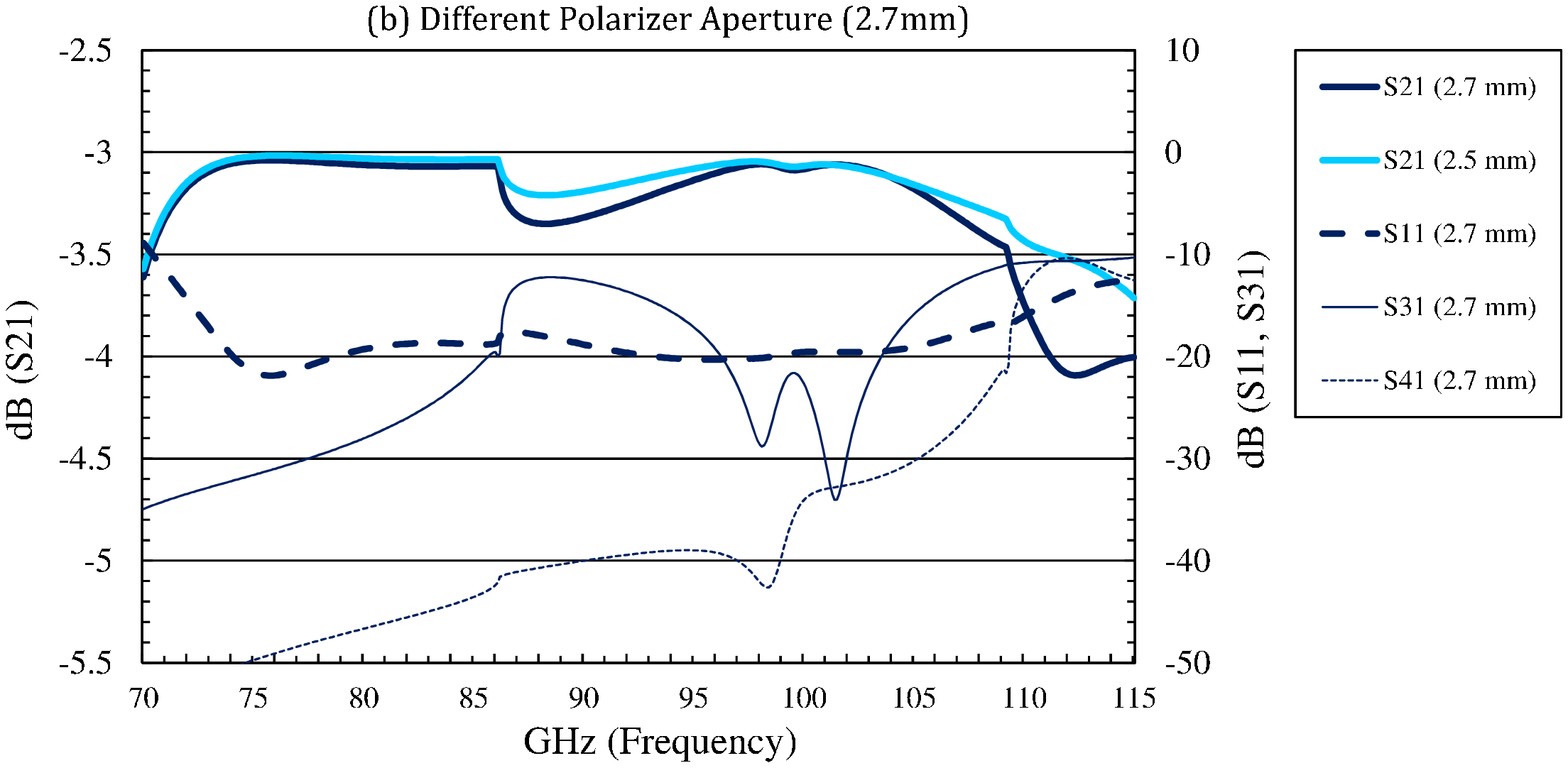}
\vspace{50pt}
\caption{Simulation results of circular polarizers with $8\%$ smaller (a) and $8\%$ larger (b) diameters.  The septa in these polarizers are the same as the one in the original polarizer of 2.5 mm diameter for a test of performance optimization.  The $S_{21}$ of Fig. (3) is rescaled in frequency and over-plotted in (a) and (b) for detailed comparisons.   The performance of both polarizers is found to be slightly worse than the original one.
\label{fig4}}
\end{figure}

\begin{figure}
\hspace{0pt} \includegraphics[angle=270,scale=.60]{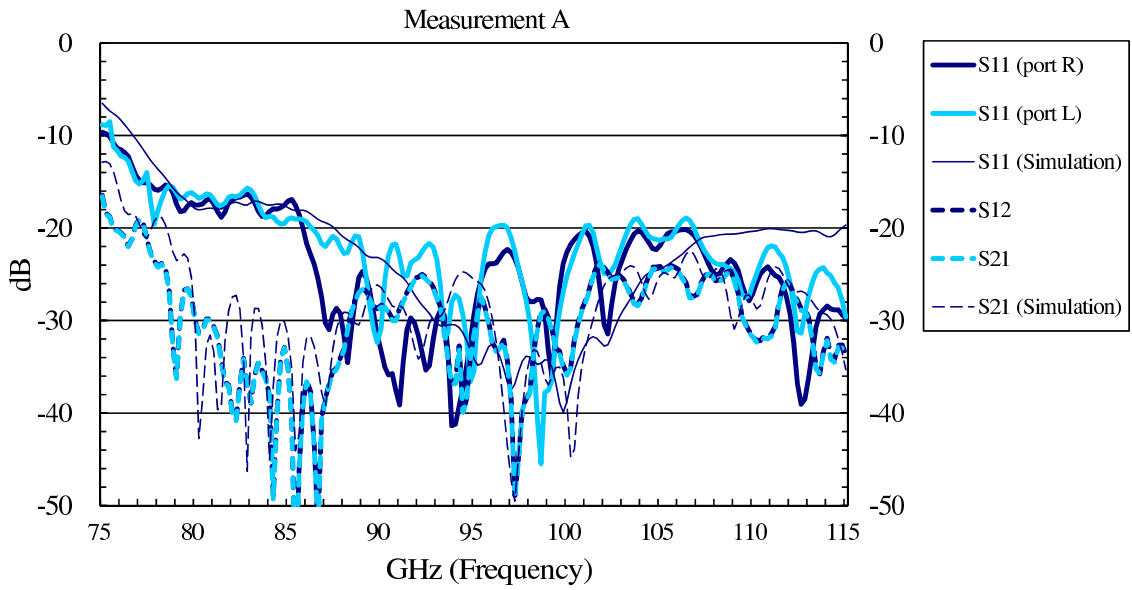}
\vspace{60pt}
\caption{Results of isolation $S_{ij}$ and reflection $S_{ii}$ for measurement A,
and results of the simulation with an identical setup as the measurement.  The overall agreement
between the two is very good.
The measured $S_{12}$ turns out to be indistinguishable from the measured
$S_{21}$.
\label{fig5}}
\end{figure}

\begin{figure}
\vspace{50pt}
\includegraphics[angle=270,scale=.50]{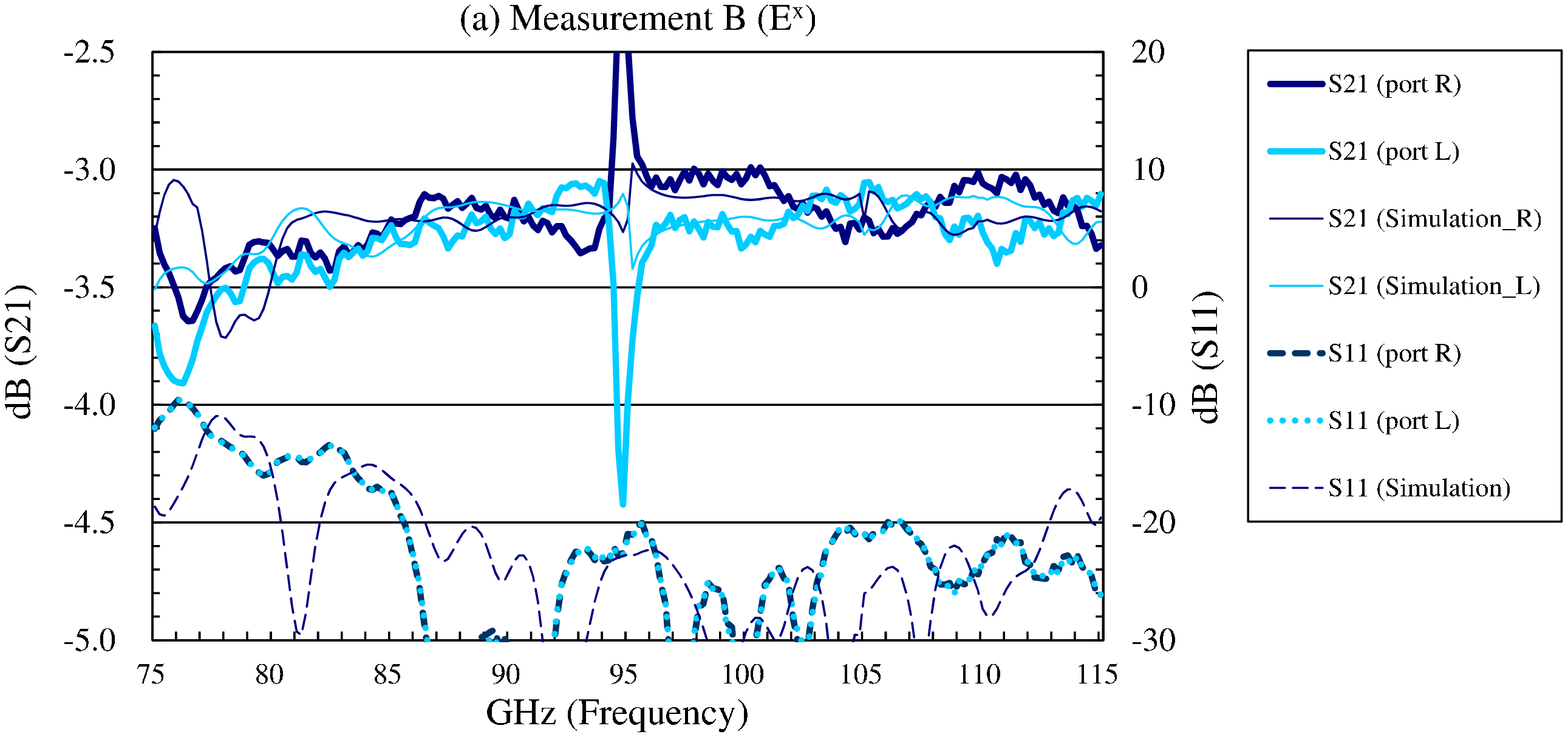} \\
\\
\includegraphics[angle=270,scale=.50]{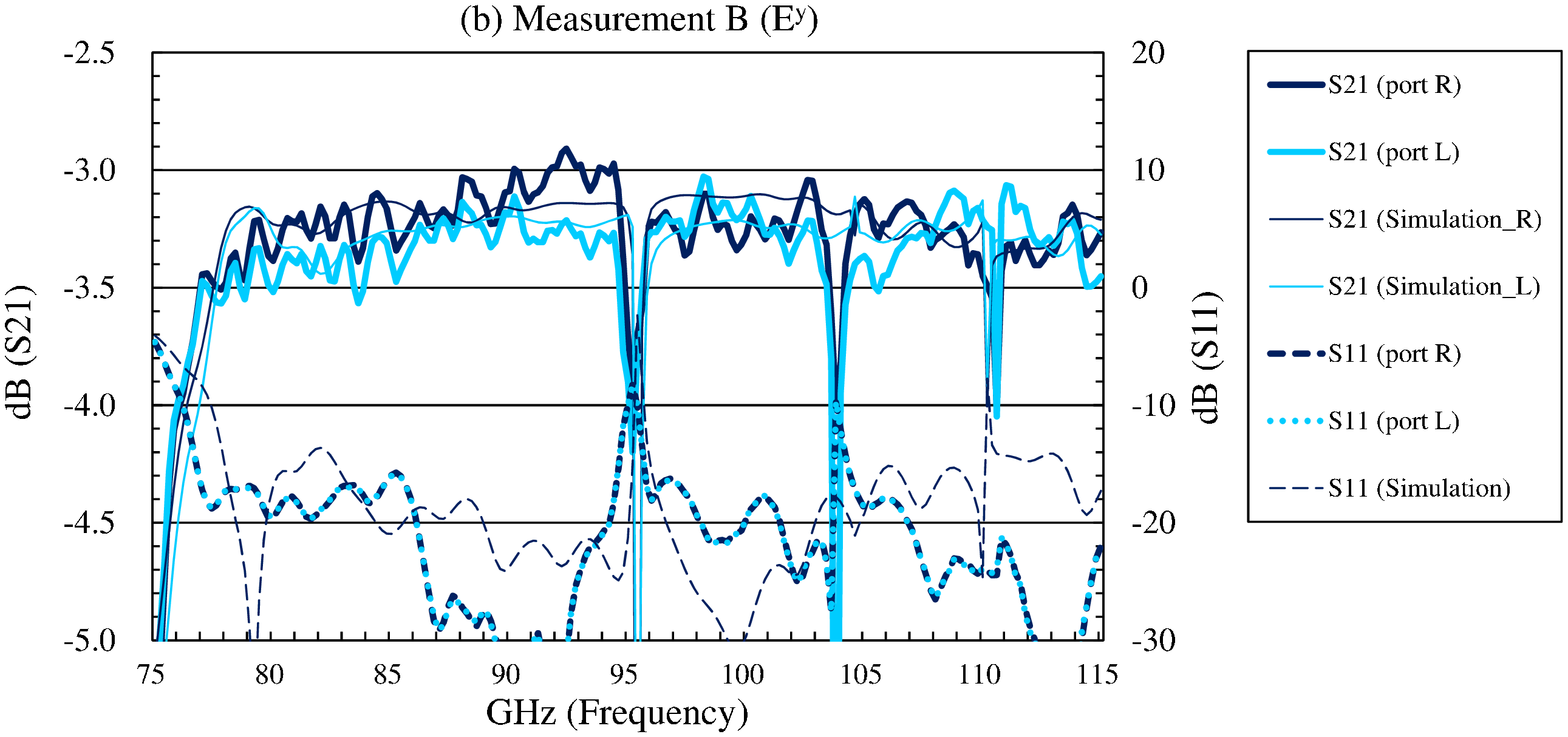}
\vspace{50pt}
\caption{The measurement B and simulation results for the $E^x$ input (a) and the $E^y$ input (b). The rectangular-to-circular transition adapter is 0.2 mm in length.
The unexpected resonance at 95 GHz shown in (a) is due to slight axis misalignment of 1.8 degrees at the interface between the adapter and the polarizer.  The agreement between measurement and simulation is considerably good; especially the measured resonances
are all captured by simulations.  Again, we find
the measured $S_{11}$'s for both $R$ and $L$ outputs are almost identical.
\label{fig6}}
\end{figure}

\begin{figure}
\includegraphics[angle=270,scale=.50]{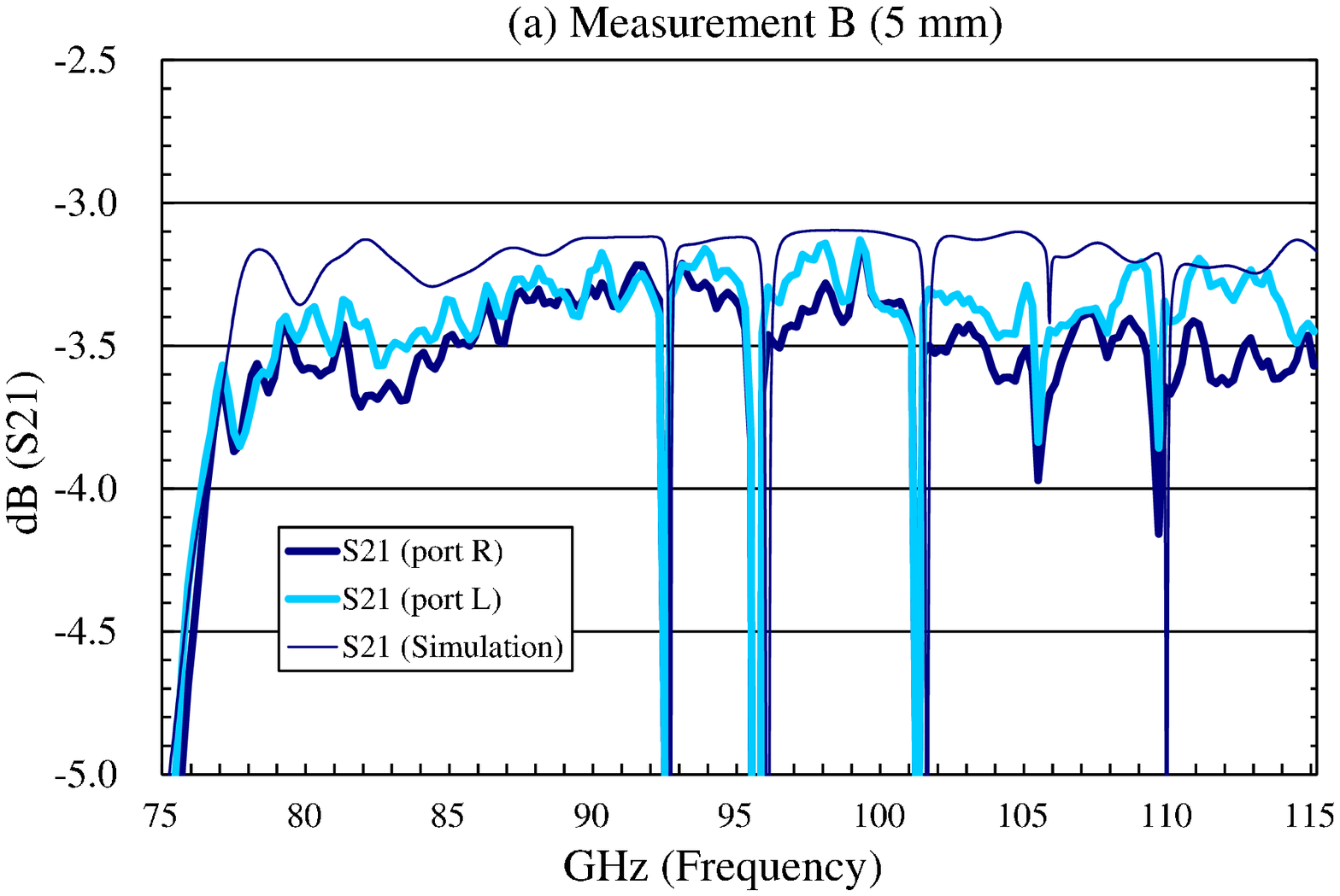} \\
\\
\\
\includegraphics[angle=270,scale=.50]{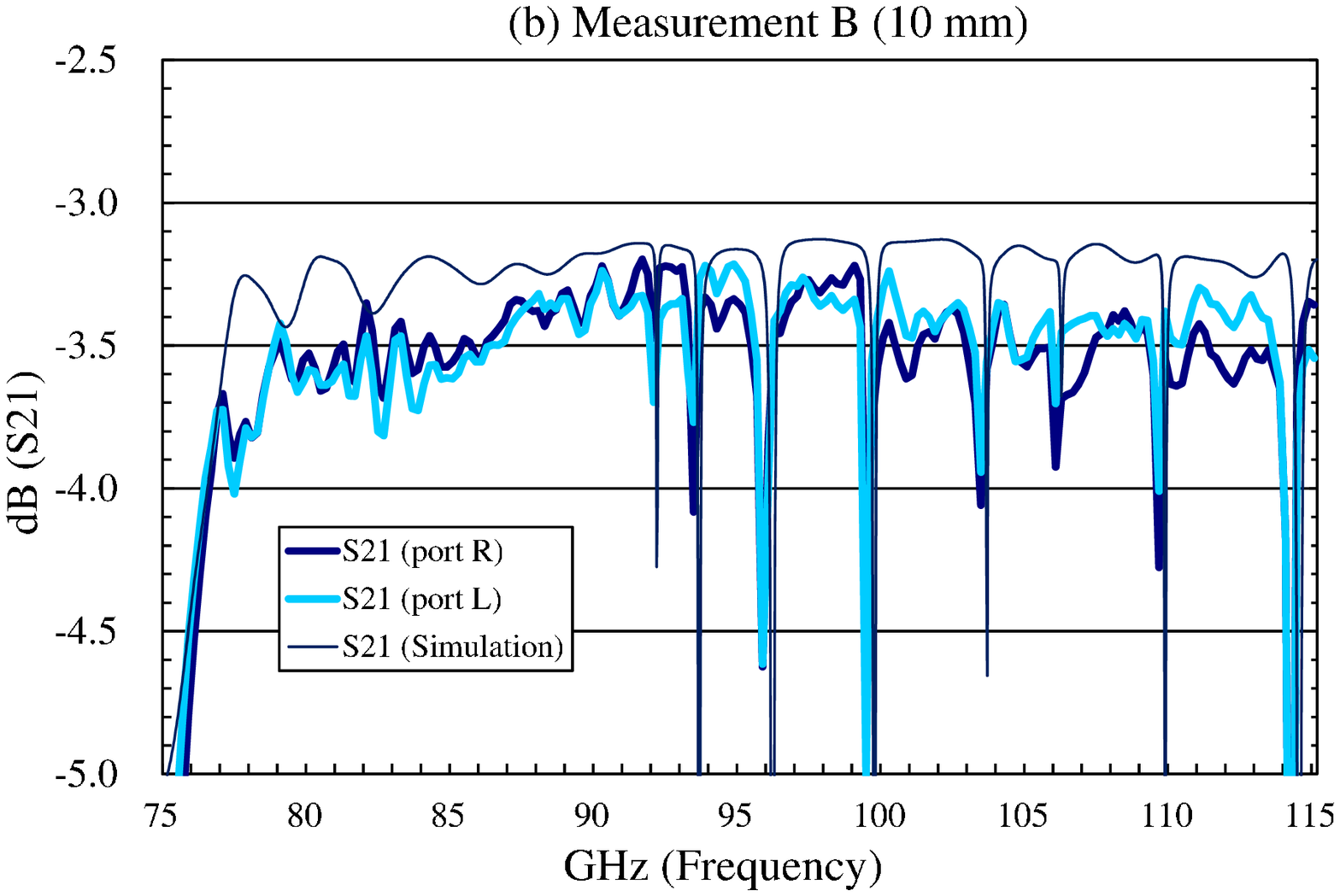} \\
\vspace{30pt}
\caption{The measurement B and simulation results for $E^y$ inputs where the original 0.2mm adapter is replaced by two other adapters of 5 mm (a)
and 10 mm (b) in length.  The measured extra losses $<0.3$ dB compared with simulation results are likely caused by the ohmic loss in the adapters.
Nevertheless we find all measured resonances are captured by simulations.
\label{fig7}}
\end{figure}

\begin{figure}
\includegraphics[angle=270,scale=.505]{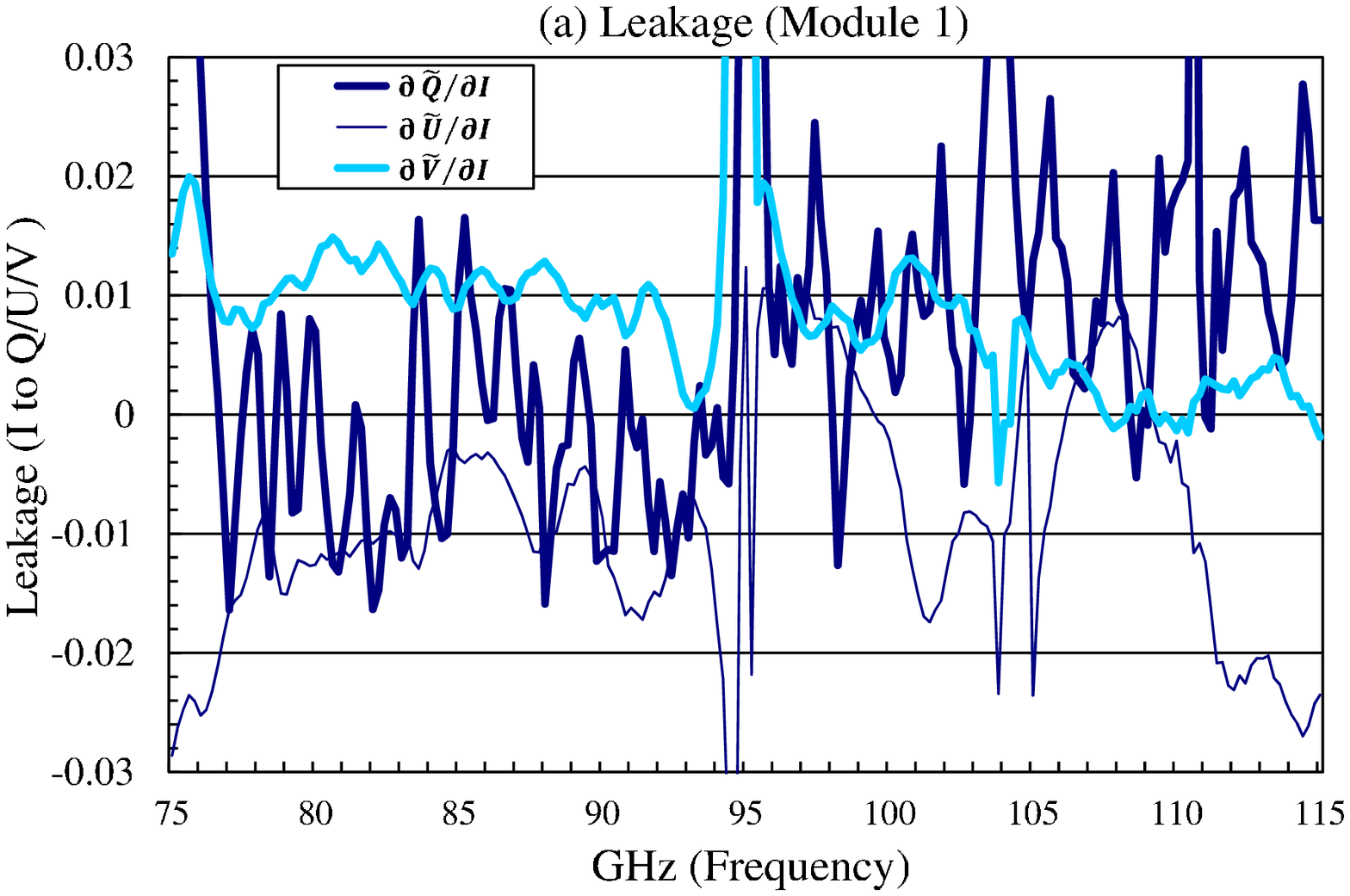} \\
\\
\\
\includegraphics[angle=270,scale=.50]{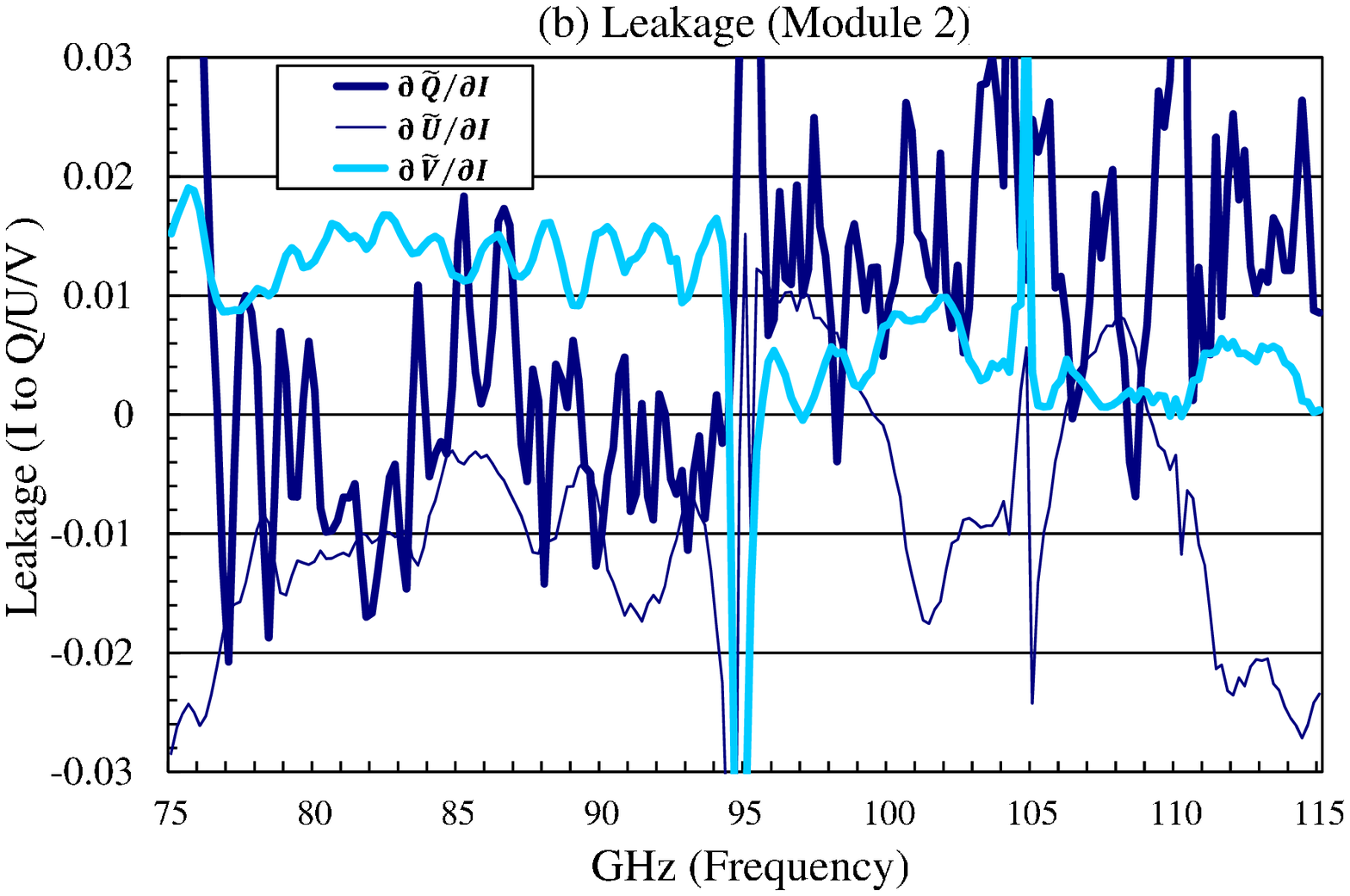} \\
\vspace{50pt}
\caption{The measured leakage from Stokes $I$ to polarized components, ($\partial\tilde Q/\partial I$,$\partial \tilde U/\partial I$, $\partial\tilde V/\partial I$ ) for the two modules (a) and (b).  The partial differentiations are taken on Eqs. (5.4), (5.5) and (5.6). \label{fig8}}
\end{figure}

\begin{figure}
\includegraphics[angle=270,scale=.50]{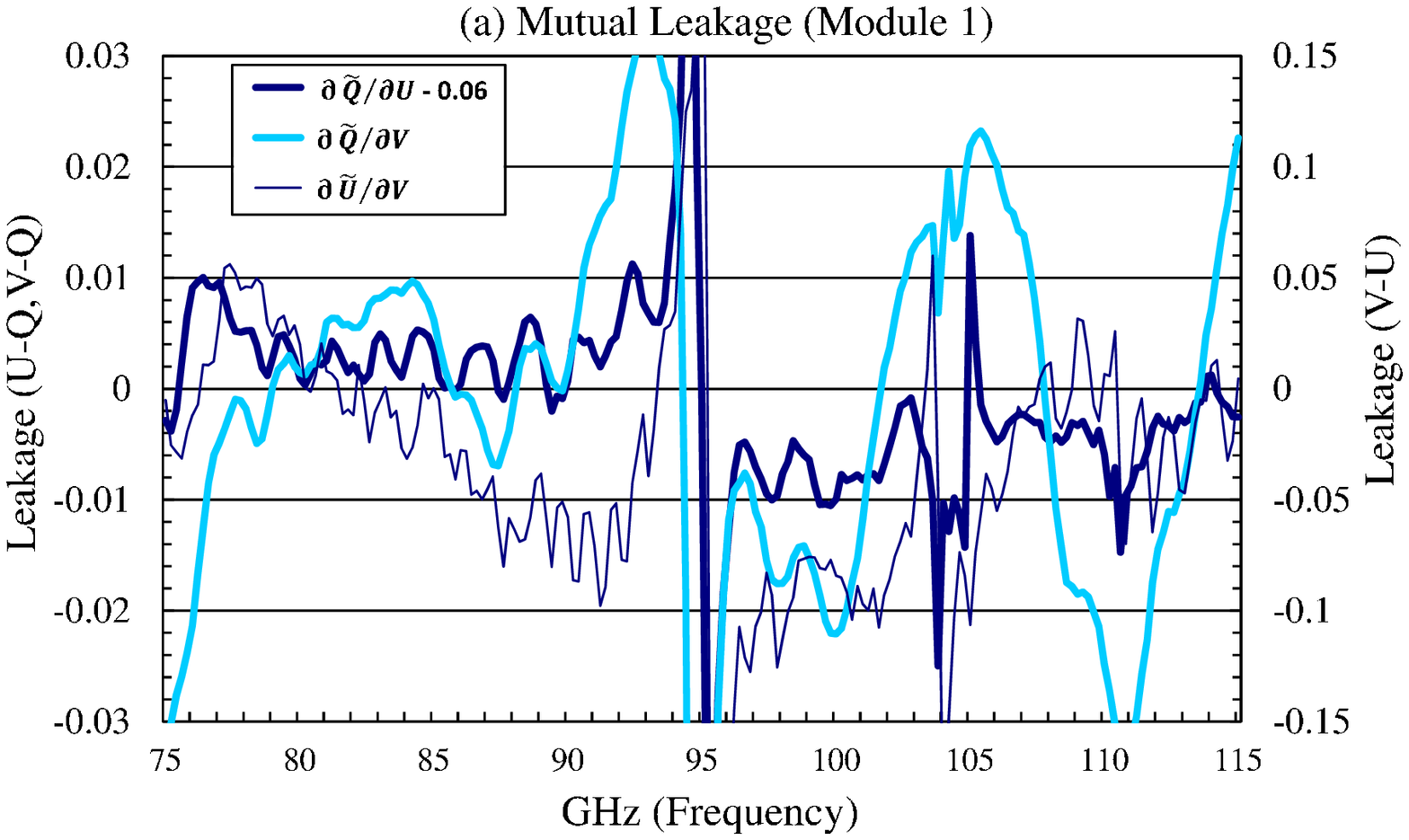} \\
\\
\\
\includegraphics[angle=270,scale=.515]{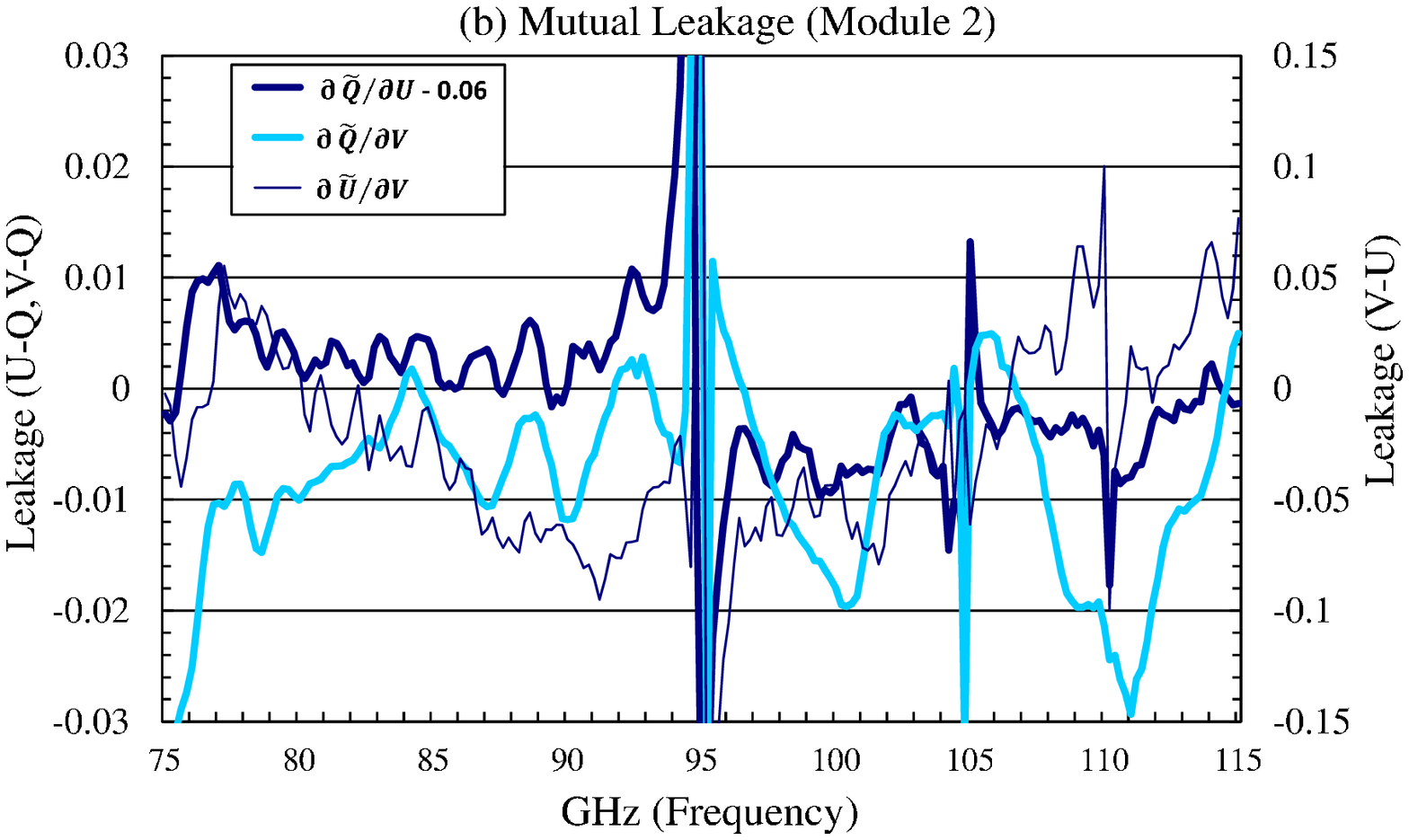} \\
\vspace{50pt}
\caption{The measured leakage among the polarized components $Q$, $U$ and $V$ for the two modules (a) and (b).  The $Q-U$ mutual leakage has a major contribution from the axis misalignment of the measurement adapter, and the resulting phase error has been corrected in this figure.  Note that the $V-U$ leakage is a few times larger than others.
\label{fig9}}
\end{figure}

\clearpage

\begin{table}
\begin{center}
\caption{Summary of Polarization Leakage \label{tbl-1}}
\vspace{20pt}
\begin{tabular}{crr}
\tableline\tableline
Leakage & 77-95 GHz & 95-115 GHz\\
\tableline
I to Q &$1\% \sim -1\%$ & $2\% \sim 0\%$\\
I to U &$0\% \sim -1\%$ & $1\% \sim -2\%$\\
I to V &$1.5\% \sim 0.5\%$ & $1\% \sim 0\%$\\
Q to U &$0.5\% \sim 0\%$ & $0\% \sim -0.5\%$\\
Q to V &$0\% \sim -1\%$ & $0\% \sim -2\%$\\
U to V &$0\% \sim -8\%$ & $5\% \sim -8\%$\\
\tableline
\end{tabular}
\end{center}
\end{table}

\end{document}